\documentstyle[12pt,a4]{article}

\newcommand{\my}{\setcounter{equation}{0}}

\begin{document}

\baselineskip=.285in

\begin{flushright}
\begin{minipage}[t]{4cm}
DPNU-96-27 \\
hep-ph/9605422 \\
May 1996
\end{minipage}
\end{flushright}

\vspace{1cm}

\begin{center}
\LARGE{
{\bf Hidden Local Symmetry \\
for Anomalous Processes \\
with Isospin/SU(3) Breaking Effects}
}

\vspace{1cm}

\large{Michio Hashimoto 
\footnote{e-mail address: michioh@eken.phys.nagoya-u.ac.jp}
} \\
\normalsize{
\it{Department of Physics Nagoya University, Nagoya 464-01 Japan }
}
\end{center}

\begin{abstract}
We show that isospin/\(SU(3)\) breaking terms can be introduced 
to the anomalous \(VVP\) coupling in the hidden local symmetry 
scheme without changing Wess-Zumino-Witten term 
in the low-energy limit. 
We make the analysis for anomalous processes of 
2-body and 3-body decays; 
radiative vector meson decays($V \to P\gamma$), 
conversion decays of photon into a lepton pair($V \to P l^+ l^-$) 
and hadronic anomalous decays($V \to PPP$). 
The predictions successfully reproduce all experimental data 
of anomalous decays. 
In particular, we predict the decay widths of 
\(\rho^0 \to \pi^0 \gamma\) and \(\phi \to \eta' \gamma \) as 
\(101\pm 9 {\rm keV \/}\) and \( 0.508 \pm 0.035 {\rm keV\/} \), 
respectively, which will be tested in the DA\(\Phi\)NE 
\(\phi\)-factory. 
Moreover, prediction is also made for \(\phi \to \pi^0 e^+ e^-\), 
\(\rho \to 3\pi\), \(K^* \to K \pi \pi\) and so on, 
for which only the experimental upper bounds are available now. 
\\

\(PACS\) number(s): 
12.39.Fe, 12.40.Vv, 13.25.-k, 13.65.+i, 14.40.Aq, 14.65.Bt
\end{abstract}

\section{Introduction}
\my

Anomalous processes involving vector mesons are interesting probes 
to test the effective theories of QCD through the low-energy and 
high-luminosity \(e^+ e^-\) collider experiments in near future.
In particular, the DA\(\Phi\)NE \(\phi\)-factory 
is expected to yield \(2\times 10^{10} \; \phi\)-meson decays per 
year [1], which will provide us with high quality data for decays of 
pseudoscalar(\(P\)) and vector mesons(\(V\)) 
in the light quark sector. 
It is expected to obtain the branching ratio of 
\(\phi\to\eta'\gamma\) [1] for which only the upper bound is 
known today [2]. Moreover, 
uncertainty of the data on \(\rho^0\to \pi^0 \gamma\) 
will be much reduced [1].

These radiative decays are associated with the flavor anomaly of QCD 
and are described by the Wess-Zumino-Witten(WZW) term[3] 
in the low energy limit. Based on the hidden local symmetry(HLS) 
[4][5][6] for the vector mesons, 
Fujiwara et al.[7] proposed 
a systematic way to incorporate vector mesons into such 
a chiral Lagrangian with WZW term without affecting 
the low-energy theorem on \(\pi^0 \to 2\gamma \; , \gamma \to 3\pi\) 
etc. Bramon et al.[8] studied extensively the radiative vector meson 
decays by introducing \(SU(3)\) breaking into the anomalous 
Lagrangian of Fujiwara et al.[7]. However, the method of Bramon et al. 
is not consistent with the low-energy theorem, 
especially on \(\eta (\eta') \to 2\gamma\), which are 
essentially determined by the WZW term. Thus, if isospin breaking 
effects were introduced through their method, successful low-energy 
theorem on \(\Gamma(\pi^0 \to 2\gamma)\) and the coupling of 
\(\gamma \to 3\pi\) would be violated. 
Furthermore, the breaking effects
(and \(\rho^0\)-\(\omega\) interference effect) 
are important to account for the difference between 
\(\Gamma(\rho^0 \to \pi^0 \gamma)\) and 
\(\Gamma(\rho^{\pm} \to \pi^{\pm}\gamma) \). 

In the previous paper[9], we proposed isospin/$SU(3)$-broken 
anomalous Lagrangians without changing the low energy theorem. 
These were obtained by eliminating direct 
\(VP\gamma\) and \(VP^3\) coupled terms, which were absent 
in the original Lagrangian[7], 
from all possible isospin/\(SU(3)\)-broken 
anomalous Lagrangians with the smallest number of derivatives. 
Then we found a parameter region 
which was consistent with all the existing data on radiative decays 
of vector mesons. 
In this paper, we give a full description of our analysis 
and \(\chi^2\)-fitting. We also include the analysis of 
$V \to P l^+ l^-$ in addition to the previous results on 
$V \to P \gamma$ and $V \to PPP$.  

The paper is organized as follows: In section 2, a review of 
HLS Lagrangian is given for both non-anomalous and anomalous terms. 
\(SU(3)\) breaking terms are introduced into the non-anomalous 
HLS Lagrangian \`a la Bando et al.[5]. 
In section 3, we construct the most general isospin/\(SU(3)\)-broken 
anomalous Lagrangians with the lowest derivatives 
in a way consistent with the low energy theorem. 
This is systematically done through spurion method for 
the breaking term. 
In section 4, the phenomenological analysis of these Lagrangians 
will be successfully done for radiative decays of vector mesons. 
In section 5, conversion decays of photon into a lepton pair 
are analyzed. 
In section 6, we make the analysis for hadronic anomalous decays. 
Section 7 is devoted to summary and discussions. 

\section{Hidden Local Symmetry}
\my

Here we give a brief review of HLS approach[6]. 
A key observation is that the non-linear sigma model based 
on the manifold \(U(3)_L \times U(3)_R/U(3)_V\) is gauge equivalent 
to another model having a symmetry 
\([U(3)_L \times U(3)_R ]_{{\rm global\/}} \times 
[U(3)_V]_{{\rm local\/}} \). Vector mesons are introduced as 
the gauge fields of a hidden local symmetry 
\([U(3)_V]_{{\rm local\/}}\). The photon field is introduced through 
gauging a part of \([U(3)_L \times U(3)_R]_{{\rm global\/}} \). 

The HLS Lagrangian is given by :[4][5] 
\begin{eqnarray}
{\cal L} &=& {\cal L}_A + a {\cal L}_V + {\cal L}_{{\rm gauge\/}}, 
\label {HLS} \\
{\cal L}_A &=& -\frac{f_{\pi}^2}{8}{\rm tr\/} (D_{\mu}\xi_L \cdot 
\xi_L^{\dagger} - D_{\mu}\xi_R \cdot \xi_R^{\dagger})^2 , 
\label{2-2} \\
{\cal L}_V &=& -\frac{f_{\pi}^2}{8} {\rm tr\/} (D_{\mu}\xi_L \cdot 
\xi_L^{\dagger}+ D_{\mu}\xi_R \cdot \xi_R^{\dagger})^2 , \label{2-3} 
\end{eqnarray}
where \(f_{\pi} = 131 {\rm MeV\/}\) is the decay constant of 
pseudoscalar mesons, \(D_{\mu} \xi_{L,R} \equiv ( 
\partial_{\mu} - igV_{\mu} )\xi_{L,R} + ie\xi_{L,R} 
Q \cdot B_{\mu}\), with 
\( Q = {\rm diag\/} \left( \frac{2}{3},-\frac{1}{3},-\frac{1}{3} 
\right) \), and with \(V_{\mu}\) and \(B_{\mu}\) 
being the vector mesons and the photon 
fields, respectively, and \({\cal L}_{{\rm gauge\/}}\) 
is the kinetic terms of \(V_{\mu}\) and \(B_{\mu}\). 
We often use an expression \(A_{\mu} \equiv Q \cdot B_{\mu}\) 
as the photon field. Here \(g\), \(e\) and \(a\) are 
respectively the hidden gauge coupling, 
the electromagnetic coupling and a free parameter not determined by 
the symmetry considerations alone. 

The fields \(\xi_{L,R}\) and \(V_{\mu}\) transform as follows; 
\begin{eqnarray}
\xi_{L,R}(x) &\to& \xi_{L,R}'(x)=h(x)\xi_{L,R}(x)
g^{\dagger}_{L,R}(x) \; , \\
V_{\mu}(x) &\to& V_{\mu}'(x)=h(x)V_{\mu}(x)h^{\dagger}(x)+ih(x)
\partial_{\mu}h^{\dagger}(x) \; , 
\end{eqnarray}
where \(h(x)\in[U(3)_V]_{{\rm local\/}}, \; 
g_{L,R}(x)\in[U(3)_{L,R}]_{{\rm global\/}}\). 
To do a phenomenological analysis, we take unitary gauge:
\begin{eqnarray}
\xi_R &=& \xi_L^{\dagger}=e^{\frac{iP}{f_\pi}} ,  \\
P &=& \left( 
\begin{array}{ccc}
\frac{\pi^0}{\sqrt{2}}+\frac{\eta}{\sqrt{3}}+\frac{\eta'}{\sqrt{6}} 
& \pi^+ & K^+ \\
\pi^- & -\frac{\pi^0}{\sqrt{2}}+\frac{\eta}{\sqrt{3}}+
\frac{\eta'}{\sqrt{6}} & K^0 \\
K^- & \bar{K}^0 & -\frac{\eta}{\sqrt{3}}+\sqrt{\frac{2}{3}}\eta'
\end{array}
\right) ,  \\
V &=& \left( 
\begin{array}{ccc}
\frac{\rho^0}{\sqrt{2}}+\frac{\omega}{\sqrt{2}} & \rho^+ & K^{*+} \\
\rho^- & -\frac{\rho^0}{\sqrt{2}}+\frac{\omega}{\sqrt{2}} & K^{*0} \\
K^{*-} & \bar{K}^{*0} & \phi
\end{array}
\right) , 
\end{eqnarray}
where we assumed that \(\eta_1\)-\(\eta_8 \) mixing angle 
\(\theta_{\eta_1-\eta_8}\) is \(\arcsin ( \frac{-1}{3}) 
\simeq -19.5\) degrees and 
\(\omega_1\)-\(\omega_8 \) mixing angle is the ideal mixing 
(35 degrees). 
If we take \(a=2\) in (\ref{HLS}), we have the celebrated 
KSRF relation \(M_{\rho}^2=2f_{\pi}^2g^2\), the universality of 
the \(\rho\)-meson coupling and the vector meson dominance for 
the electromagnetic form factor(VMD) [4]. 

For obtaining the pseudoscalar meson mass terms, we introduce 
the quark mass matrix(\({\cal M}\)) as, 
\begin{equation}
{\cal L}_M = \frac{f_{\pi}^2 \mu}{2}{\rm tr\/}(\xi_R{\cal M}
\xi_L^{\dagger}+\xi_L{\cal M}\xi_R^{\dagger}) + m_{\eta_1}^2 , 
\end{equation} 
where \(\mu {\cal M}\) is related to the mass of 
\(\pi\), \(K\) and \(\eta\), and 
\(m_{\eta_1}\) is the mass of \(\eta'\) due to \(U(1)_A\) 
breaking by the gluon anomaly. 
Analogously, we may add appropriate \(SU(3)\) breaking 
terms to (\ref{HLS}) [5], 
\begin{eqnarray}
\Delta{\cal L}_{A,(V)} &=& -\frac{f_{\pi}^2}{8}{\rm tr\/} 
(D_{\mu}\xi_L \cdot \epsilon_{A,(V)}\xi_R^{\dagger} 
\pm D_{\mu}\xi_R \cdot \epsilon_{A,(V)}\xi_L^{\dagger})^2 , \\
\epsilon_{A,(V)} &=& {\rm diag\/}(0,0,\epsilon_{A,(V)}).
\end{eqnarray}
Even if those $SU(3)$ breaking terms are introduced, 
we can show the successful relations [5]:
\begin{equation}
\frac{g_{\rho}}{M_{\rho}^2}=
\frac{3g_{\omega}}{M_{\omega}^2}=
-\frac{3g_{\phi}}{\sqrt{2}M_{\phi}^2}=\frac{1}{g} .
\label{2-12}
\end{equation}
We will use this relations, when we consider radiative decays of 
vector mesons and conversion decays of photon into a lepton pair. 

Further improvements of (\(\ref{HLS}\)) have been elaborated in 
Ref.[10]. 
Here we will not discuss the non-anomalous sector (\(\ref{HLS}\)) 
any furthermore, 
because we are only interested in the anomalous sector. 
We simply assume that the parameters of the non-anomalous 
Lagrangian have been arranged so as to reproduce the relevant 
experimental data. Thus we use the experimental values as inputs 
from the non-anomalous part. 

In addition to (\(\ref{HLS}\)) there exists an anomalous part of 
the HLS Lagrangian. Fujiwara et al.[7] proposed how to incorporate 
vector mesons into this part of the Lagrangian without changing 
the anomaly determined by WZW term[7]. 
They have given the anomalous action as follows:
\begin{equation}
\Gamma = \Gamma_{WZW} + \sum_{i=1}^4 \int_{M^4}c_i{\cal L}_i , 
\label{WZ} 
\end{equation}
where 
\begin{eqnarray}
\Gamma_{WZW} &=& -\frac{iN_c}{240\pi^2} \int_{M^5}{\rm tr\/}
[(dU)\cdot U^{\dagger}]^5_{{\rm covariantization\/}} \; ,  \\
{\cal L}_1 &=& {\rm tr\/} (\hat{\alpha}_L^3\hat{\alpha}_R - 
\hat{\alpha}_R^3\hat{\alpha}_L) ,  \\
{\cal L}_2 &=&{\rm tr\/} (\hat{\alpha}_L\hat{\alpha}_R\hat{\alpha}_L 
\hat{\alpha}_R) ,  \\
{\cal L}_3 &=& i {\rm tr\/} F_V(\hat{\alpha}_L \hat{\alpha}_R 
- \hat{\alpha}_R\hat{\alpha}_L) ,  \\
{\cal L}_4 &=& \frac{i}{2} {\rm tr\/} (\hat{F}_L+\hat{F}_R)\cdot
(\hat{\alpha}_L\hat{\alpha}_R-\hat{\alpha}_R\hat{\alpha}_L) , \\
\hat{\alpha}_{L,R} &=& D\xi_{L,R}\cdot\xi_{L,R}^{\dagger}=
d\xi_{L,R}\cdot \xi_{L,R}^{\dagger}-igV+ie\xi_{L,R}A
\xi_{L,R}^{\dagger} \; ,  \\
U &=& \xi_L^{\dagger}\xi_R \; ,\qquad  F_V = dV-igV^2,  \\
\hat{F}_{L,R} &=& \xi_{L,R}(dA - ieA^2)\xi_{L,R}^{\dagger} \: .  
\end{eqnarray}
Notice that \({\cal L}_1 \sim {\cal L}_4 \) have no contribution to 
anomalous processes such as \(\pi^0\to 2\gamma\) and 
\(\gamma\to 3\pi\) at soft momentum limit, 
because these Lagrangian are constructed with hidden-gauge covariant 
blocks such as \(\hat{\alpha}_{L,R},F_V \) and \(\hat{F}_{L,R} \)[7]. 

We take \(c_3=c_4=-15C , c_1-c_2=15C \) in (\ref{WZ}) 
for phenomenological reason[7]. 
Then we obtained the Lagrangian of anomalous sector as follows:
\begin{eqnarray}
{\cal L}_{{\rm FKTUY\/}} & = & \frac{-iN_c}{48 \pi^2} 
\left[ 3(VVP)-2(\gamma P^3) \right] + \cdot \cdot \cdot , 
\label{VVP} \\
(VVP) &=& -\frac{2ig^2}{f_{\pi}} {\rm tr\/} (VdVdP+dVVdP) , 
\nonumber \\
(\gamma P^3) &=& \frac{4e}{f_{\pi}^3}{\rm tr\/}A(dP)^3 . 
\nonumber 
\end{eqnarray}

Here, it is important that the amplitude such as 
\(\pi^0 \to 2\gamma, \gamma \to 3\pi\) at low energy limit are 
determined only by the non-Abelian anomaly of 
the chiral \(U(3)_L  \times U(3)_R \) symmetry. 
The Lagrangian \({\cal L}_{\rm FKTUY\/}\) is, 
of course, consistent with the low energy theorem related to 
the anomaly. 

\section{Isospin/\(SU(3)\)-breaking Terms in the Anomalous Sector}
\my

We now consider how to modify \({\cal L}_1 \sim {\cal L}_4\) 
by introducing isospin/\(SU(3)\)-breaking parameters, 
\(\epsilon\)'s, treated as ``spurions''[11]. 
The spurion \(\epsilon\) transforms as 
\(\epsilon \to g_L \, \epsilon \, g_R^{\dagger}\). 
Then we define the hidden-gauge covariant block 
\(\hat{\epsilon} \equiv \frac{1}{2}(\xi_L \epsilon \xi_R^{\dagger} + 
\xi_R \epsilon^{\dagger} \xi_L^{\dagger})\). 
We construct Lagrangians out of the hidden-gauge covariant blocks 
such as \(\hat{\alpha}_{L,R}\), \(F_V\), \(\hat{F}_{L,R}\) 
and \(\hat{\epsilon}\) so as to make them `` invariant " under 
\([U(3)_L  \times U(3)_R]_{{\rm global\/}} \times 
[U(3)_V]_{{\rm local\/}} \) 
as well as parity(\(P\))-, charge conjugation(\(C\))- and 
\(CP\)-transformations. 
After hidden-gauge fixing, they become explicit breaking terms of 
the \(SU(3)\) symmetry. 
Then, in general, we obtain isospin/\(SU(3)\)-broken anomalous 
Lagrangians with the lowest number of derivatives.
\begin{eqnarray}
\Delta {\cal L}_1 &=& {\rm tr\/} [\hat{\alpha}_L^3( \hat{\alpha}_R 
\cdot \hat{\epsilon}^{(1)} + \hat{\epsilon}^{(1)} \cdot 
\hat{\alpha}_R) - \hat{\alpha}_R^3(\hat{\alpha}_L \cdot 
\hat{\epsilon}^{(1)} + \hat{\epsilon}^{(1)} \cdot \hat{\alpha}_L)]  
, \\
\Delta {\cal L}'_1 &=& {\rm tr\/} (\hat{\alpha}_L 
\hat{\epsilon}^{(1')}\hat{\alpha}_L^2 \hat{\alpha}_R - 
\hat{\alpha}_R \hat{\epsilon}^{(1')}\hat{\alpha}_R^2 \hat{\alpha}_L 
+ \hat{\alpha}_L^2 \hat{\epsilon}^{(1')} \hat{\alpha}_L 
\hat{\alpha}_R   -\hat{\alpha}_R^2 \hat{\epsilon}^{(1')} 
\hat{\alpha}_R \hat{\alpha}_L),  \\
\Delta {\cal L}_2 &=&{\rm tr\/} (\hat{\epsilon}^{(2)} \cdot 
\hat{\alpha}_L + \hat{\alpha}_L \cdot \hat{\epsilon}^{(2)}) 
\hat{\alpha}_R\hat{\alpha}_L\hat{\alpha}_R , \\
\Delta {\cal L}_3 &=& i {\rm tr\/} (F_V \cdot \hat{\epsilon}^{(3)} + 
\hat{\epsilon}^{(3)} \cdot F_V) \cdot (\hat{\alpha}_L \hat{\alpha}_R 
- \hat{\alpha}_R\hat{\alpha}_L)  , \\
\Delta {\cal L}'_3 &=& i{\rm tr\/} F_V (\hat{\alpha}_L 
\hat{\epsilon}^{(3')} \hat{\alpha}_R - \hat{\alpha}_R 
\hat{\epsilon}^{(3')} \hat{\alpha}_L),  \\
\Delta {\cal L}_4 &=& i {\rm tr\/} [\{ \: (\hat{F}_L+\hat{F}_R) \cdot 
\hat{\epsilon}^{(4)} + \hat{\epsilon}^{(4)} \cdot (\hat{F}_L+\hat{F}_R)
 \: \} \cdot 
(\hat{\alpha}_L\hat{\alpha}_R -\hat{\alpha}_R \hat{\alpha}_L)], \\
\Delta {\cal L}'_4 &=& i{\rm tr\/} (\hat{F}_L+\hat{F}_R)\cdot 
(\hat{\alpha}_L \hat{\epsilon}^{(4')}
\hat{\alpha}_R - \hat{\alpha}_R \hat{\epsilon}^{(4')}\hat{\alpha}_L) 
, \\
\Delta{\cal L}_5 &=& {\rm tr\/} \hat{\epsilon}^{(5)}
(\hat{\alpha}_L^2  \hat{\alpha}_R^2 - 
\hat{\alpha}_R^2 \hat{\alpha}_L^2) ,  \\
\Delta{\cal L}_6 &=& i {\rm tr\/} (\hat{\epsilon}^{(6)}F_V - F_V 
\hat{\epsilon}^{(6)})\cdot (\hat{\alpha}_L^2- \hat{\alpha}_R^2) , \\
\Delta{\cal L}_7 &=& i {\rm tr\/} [(\hat{\epsilon}^{(7)}\hat{F}_L 
- \hat{F}_L \hat{\epsilon}^{(7)})
\hat{\alpha}_R^2 -(\hat{\epsilon}^{(7)}\hat{F}_R - \hat{F}_R 
\hat{\epsilon}^{(7)})\hat{\alpha}_L^2] , \\
\Delta{\cal L}_8 &=& i {\rm tr\/}[(\hat{\epsilon}^{(8)}\hat{F}_L 
- \hat{F}_L \hat{\epsilon}^{(8)})
 \hat{\alpha}_L^2-(\hat{\epsilon}^{(8)}\hat{F}_R - \hat{F}_R 
 \hat{\epsilon}^{(8)})\hat{\alpha}_R^2] . 
\end{eqnarray}
Here \(\hat{\alpha}_{L,R}\), \(F_V\), \(\hat{F}_{L,R}\) transform 
under \(P\) and \(C\) transformations as 
\begin{eqnarray}
P: \; \; & & \hat{\alpha}_{L,R\mu}\longrightarrow 
\hat{\alpha}_{R,L}^{\mu} \; ,\\
& & F_{V\mu\nu}\longrightarrow F_V^{\mu\nu} \: , \: 
\hat{F}_{L,R\mu\nu}\longrightarrow \hat{F}_{R,L}^{\mu\nu} \; , \\
C: \; \; & & \hat{\alpha}_{L,R}\longrightarrow -\hat{\alpha}_{R,L}^T 
\; , \\
& & F_V \longrightarrow -F_V^T \: , \: 
\hat{F}_{L,R} \longrightarrow -\hat{F}_{R,L}^T \: .
\end{eqnarray}
We could introduce another ``spurion'' \(\hat{\epsilon}_- = 
\frac{1}{2}(\xi_L \epsilon \xi_R^{\dagger} - 
\xi_R \epsilon^{\dagger} \xi_L^{\dagger})\), which, however, 
is not relevant to the following analysis. 

There still exist too many parameters. However, 
we may select the combination of \(\Delta{\cal L}_{1 \sim 8} \) 
so as to eliminate 
the direct \(V\gamma P\)-, \(VP^3\)-coupling terms, 
which do not exist in the original 
Lagrangian \({\cal L}_{{\rm FKTUY\/}}\). 
Then the isospin/\(SU(3)\)-broken anomalous Lagrangians consist of 
only the following two terms: 
\begin{eqnarray}
-\Delta {\cal L}_{VVP}^a &=& \frac{3g^2}{4\pi^2 f_P}
{\rm tr\/}\epsilon'(dVdVP+PdVdV) - \frac{3e^2}{4\pi^2 f_P}{\rm tr\/}
\epsilon'(dAdAP+PdAdA) \nonumber \\   
 & & + i \frac{3e}{4\pi^2 f_P^3}{\rm tr\/}\epsilon'
 (dP^3A-AdP^3+dPAdP^2-dP^2AdP), \\
-\Delta {\cal L}_{VVP}^b &=& \frac{3g^2}{2\pi^2 f_P}{\rm tr\/}
\epsilon(dVPdV) 
- \frac{3e^2}{2\pi^2 f_P}{\rm tr\/}\epsilon(dAPdA) \nonumber \\
 & & + i \frac{3e}{2\pi^2 f_P^3} {\rm tr\/}\epsilon(dP^3A-AdP^3) .  
\end{eqnarray}
We can also understand these \(\Delta {\cal L}_{VVP}^{a,b} \) 
in a more straightforward way: We can introduce the breaking terms 
\(\epsilon\) to the first \((VVP)\)-term 
in the original \({\cal L}_{\rm FKTUY\/}\) via two possible ways, 
which correspond to the first terms of 
\(\Delta {\cal L}_{VVP}^{a,b} \). Next, we determine 
\(\gamma \gamma P\)-, \(\gamma P^3\)-terms 
so as to eliminate them at soft momentum 
limit by using the relation 
\(gV \to eA + \frac{i}{2f_{\pi}^2}[P,dP] \) 
as well as to make them invariant under \(P\), \(C\) 
and \(CP\)-transformations. These terms correspond to 
the second and third terms of \(\Delta {\cal L}_{VVP}^{a,b} \). 

Our \(\Delta {\cal L}_{VVP}^b \) resembles the \(SU(3)\)-broken 
anomalous Lagrangian introduced by Bramon et al.[8], 
but is conceptually quite different from the latter. In fact 
the prediction on \(\eta (\eta') \to 2\gamma \) decay width 
in the latter disagrees with the low energy theorem. 
On the other hand, our \(\Delta{\cal L}_{VVP}^{a,b}\) obviously 
do not change the low energy theorem by construction. 

\section{Phenomenological Analysis for Radiative Decays}
\my
We now discuss phenomenological consequences of our Lagrangian 
\({\cal L}_{{\rm anomalous\/}}={\cal L}_{{\rm FKTUY\/}} + 
\Delta {\cal L}_{VVP}^a + \Delta {\cal L}_{VVP}^b \). 
For convenience, we define relevant coupling  constant as
\begin{equation}
g_{VP\gamma}=\sum_{V'}\frac{g_{VV'P}g_{V'}}{M_{V'}^2},
\end{equation}
considering that these decays proceed via intermediate vector mesons 
\(V'\). Then we obtain each radiative decay width
\begin{eqnarray}
\Gamma(V\longrightarrow P\gamma) &=& \frac{1}{3}\alpha \cdot 
g_{VP\gamma}^2 \left( \frac{M_V^2-M_P^2}{2M_V} \right) ^3 , \\
\Gamma(\eta'\longrightarrow V\gamma) &=& \alpha \cdot 
g_{\eta' V\gamma}^2 
\left(\frac{M_{\eta'}^2 - M_V^2}{2M_{\eta'}} \right)^3 , 
\end{eqnarray} 
where \(g_{VV'P}\), \(g_{V'}\) and \(M_{V'}\) are anomalous 
\(VV'P\) coupling constant, \(V'\)-\(\gamma\) mixing 
and mass of the vector meson, respectively. 
In \(\Delta{\cal L}_{VVP}^{a,b}\) we take a parametrization 
for convenience:
\begin{equation}
\epsilon' = \left( 
\begin{array}{ccc}
-\epsilon'_1 &  &  \\
 & -\epsilon'_2 &  \\
 &  & -\epsilon'_3
\end{array}  
\right) , \;\;
\epsilon = \left( 
\begin{array}{ccc}
\epsilon_1 + \epsilon'_1 &  &  \\
 & \epsilon_2+\epsilon'_2 &  \\
 &  & \epsilon_3+\epsilon'_3
\end{array}  
\right). \label{4-4}
\end{equation}
Thus each \(g_{VP\gamma}\) is given in terms of the parameters in 
\(\Delta{\cal L}_{VVP}^{a,b}\) :
\begin{equation}
\left \{
\begin{array}{rcl}
g_{\rho^0\pi^0\gamma} &=& G(1+4\epsilon_1-2\epsilon_2+3\delta) ,\\
g_{\rho^{\pm}\pi^{\pm}\gamma} &=& G(1+3\epsilon'_1-3\epsilon'_2 
+4\epsilon_1-2\epsilon_2) , \\
g_{\omega\pi^0\gamma} &=& 3G(1+\frac{4}{3}\epsilon_1+
\frac{2}{3}\epsilon_2-\frac{\delta}{3}) ,\\
g_{\omega\eta\gamma}&=& \frac{f_{\pi}}{f_{\eta}}\sqrt{\frac{2}{3}}G
(1+4\epsilon_1-2\epsilon_2-\sqrt{2}\theta_V-3\delta-
\frac{\theta_P}{\sqrt{2}}) , \\
g_{\rho^0\eta\gamma} &=& \frac{f_{\pi}}{f_{\eta}}\sqrt{6}G
(1+\frac{4}{3}\epsilon_1+\frac{2}{3}\epsilon_2+\frac{\delta}{3}-
\frac{\theta_P}{\sqrt{2}}), \\
g_{\phi\eta\gamma}&=& \frac{f_{\pi}}{f_{\eta}}\frac{2}{\sqrt{3}}G
(1+2\epsilon_3+\frac{\theta_V}{\sqrt{2}}+\sqrt{2}\theta_P) , \\
g_{K^{*\pm}K^{\pm}\gamma} &=& \frac{f_{\pi}}{f_K}G(1+3\epsilon'_1
-3\epsilon'_3
+4\epsilon_1-2\epsilon_3) ,\\ 
g_{\bar{K^{*0}}\bar{K^0}\gamma} &=& -\frac{f_{\pi}}{f_K}2G
(1+\epsilon_2+\epsilon_3) , \\
g_{\phi\pi^0\gamma} &=& g_{\omega\pi^0\gamma} \cdot\theta_V , \\
g_{\eta'\rho^0\gamma} &=& \frac{f_{\pi}}{f_{\eta'}}\sqrt{3}G
(1+\frac{4}{3}\epsilon_1+\frac{2}{3}\epsilon_2+
\frac{\delta}{3}+\sqrt{2}\theta_P) , \\
g_{\eta'\omega  \gamma} &=& \frac{f_{\pi}}{f_{\eta'}}
\frac{1}{\sqrt{3}}G
(1+4\epsilon_1-2\epsilon_2+2\sqrt{2}\theta_V-
3\delta+\sqrt{2}\theta_P) , \\
g_{\phi\eta'\gamma} &=& -\frac{f_{\pi}}{f_{\eta'}}
\frac{2\sqrt{2}}{\sqrt{3}}G(1+2\epsilon_3-\frac{\theta_V}{2\sqrt{2}}
-\frac{\theta_P}{\sqrt{2}}) , \label{4-5}
\end{array}
\right.
\end{equation}
where \(G = \frac{g}{4\pi^2f_{\pi}}\) and 
we used the relations of (\ref{2-12}). 

The parameters \(\theta_V,\theta_P\) 
appearing in the expression of \(g_{VP\gamma}\) stand for 
the deviation of \(\phi\)-\(\omega\), \(\eta\)-\(\eta'\) 
mixing angles from the ideal mixing and 
\(\eta_1\)-\(\eta_8\) mixing, respectively. 
The parameter \(\delta\) comes from 
the \(\rho\)-\(\omega\) interference effect arising from 
the small mass difference of \(\rho\) and \(\omega\). 

For reproducing the experimental value of 
\(\Gamma (\phi\to\rho\pi\to\pi\pi\pi)\), 
we took \(\theta_V=0.0600\pm 0.0017 \). The sign comes from 
the observed \(\phi\)-\(\omega\) interference effects in 
\(e^+e^- \to \pi^+ \pi^- \pi^0 \)[2]. 

The mixing angle \(\theta_{\eta_1-\eta_8}( =\)arcsin(\(-\)1/3)) 
has been supported in \(\eta\)-\(\eta'\) phenomenology[13], 
thus, it is admitted to take \(\theta_P=0\). 

Similarly, we consider the decay of 
\(\omega\to\pi\pi\), which is \(G\)-parity violating process. 
If the isospin were not broken, such process would not exist. 
The experimental value 
of \(\Gamma(\omega \to \pi \pi) \) is reproduced for 
\(\delta=0.0348\pm 0.0024\). 
We calculated \(\delta\) from \(\Gamma(\omega\to\pi\pi)/ 
\Gamma(\rho\to\pi\pi)=
\delta^2 \cdot \frac{p_{\omega\to\pi\pi}^3}{M_{\omega}^2}/
\frac{p_{\rho\to\pi\pi}^3}{M_{\rho}^2}\), where \(p_{V\to\pi\pi}\) 
is the final state pion momentum. 
The ambiguity of the sign has been resolved recently 
through the decays of \(\omega\) produced in 
\(\pi^- p \to \omega n\) [12], in which the constructive interference 
has been supported. 

There are essentially five free parameters from 
\(\Delta {\cal L}_{VVP}^{a,b}\) in (\ref{4-5}), 
because \(\epsilon'_1\) is negligible. 
We determine these parameters by \(\chi^2\) fitting, 
using the data of the radiative vector meson decays 
and \(\omega \to 3 \pi\). Then we obtain 
$\epsilon_1=-0.0174 \pm 0.0100$, 
$\epsilon_2=-0.0246 \pm 0.0114$, $\epsilon_3=-0.0974 \pm 0.0103$, 
$\epsilon'_1 - \epsilon'_2 = -0.0292 \pm 0.0015 $ and 
$\epsilon'_1 - \epsilon'_3 = 0.0366 \pm 0.0028 $.

We take \(g=4.27\pm 0.02\) from \(\Gamma(\rho \to \pi \pi)=151.2\pm 
1.2 {\rm MeV\/}\), and \(f_{\pi}=131{\rm MeV\/},f_K=160 \pm 2
{\rm MeV\/}\)[2], and \(f_{\eta}=150\pm 6{\rm MeV\/},
f_{\eta'}=142\pm 3{\rm MeV\/}\) from 
\(\eta (\eta') \to 2\gamma\) [2]. 
Then we obtained the results listed in Table I. 

In Table I, (i)\(\sim\)(iii) mean:

\(\hspace{-1cm} \left \{
\begin{array}{l}
\mbox{\enskip (i) Values of original } \; {\cal L}_{{\rm FKTUY\/}} 
. \\
\mbox{ (ii) Values of the } SU(3)
\mbox{ -broken model by Bramon et al.[8] } 
(\epsilon_3 = -0.1 \pm 0.03). \\
\mbox{(iii) Values of our model. } \\
\qquad  \mbox{ The parameter values are } 
\epsilon_1=-0.0174 \pm 0.0100, 
\epsilon_2=-0.0246 \pm 0.0114, \\
\qquad \epsilon_3=-0.0974 \pm 0.0103, 
\epsilon'_1 - \epsilon'_2 = -0.0292 \pm 0.0015 \mbox{ and } 
\epsilon'_1 - \epsilon'_3 = 0.0366 \pm 0.0028 .
\end{array}
\right.
\)
\\

These parameters suggest that isospin/\(SU(3)\)-breaking 
effects for the anomalous sector cannot be given by the quark mass 
matrix in a simple manner. We discuss this point later. 

In the previous paper[9], we determined a parameter region so as to 
reproduce all experimental data of radiative vector meson decays as 
\(0.0279 < 4 \epsilon_1 - 2 \epsilon_2 < 0.0670 \), 
\(-0.0471< \frac{4}{3}\epsilon_1 +\frac{2}{3}\epsilon_2 <-0.0174\), 
\(-0.0112 < \epsilon_2 + \epsilon_3 < -0.0902 \), 
\(-0.0925 < \epsilon_3 < -0.0702\), 
\(4\epsilon_1 -2\epsilon_2 +3\epsilon'_1-3\epsilon'_2=-0.108\) and 
\(4\epsilon_1 - 2\epsilon_3 + 3\epsilon'_1-3\epsilon'_3=0.235 \). 
The parameters by \(\chi^2\)-fitting are slightly different 
from the above region,  which then yield the prediction of 
\(\Gamma(\omega \to \eta \gamma) \) from the experimental value. 
However, the difference of the former maximum value 
from the latter minimum value is about $5 \%$. 
If we consider that the experimental data[2] is determined only 
by Ref.[12], where large momentum transfer events have been selected 
in order to eliminate \(\rho\)-\(\omega\) interference contribution, 
our result is not inconsistent with the experiments. 
In fact, if we take the experimental 
value of Ref.[14], where the decay width of 
\(\omega \to \eta \gamma\) has been reported as \(6.2 \pm 2.4 \)keV 
based on the assumption of constructive \(\rho\)-\(\omega\) 
interference, our prediction by \(\chi^2\)-fitting also reproduces 
the experimental value. 

The results for \(\Gamma(\rho^0\to\pi^0\gamma)\), 
\(\Gamma(\rho^{\pm}\to\pi^{\pm}\gamma)\) and 
\(\Gamma(\omega\to\pi^0\gamma)\) in Table I 
suggest that isospin breaking terms are very important. 
Both (i) and (ii) in Table I do not have isospin breaking terms. 
These values differ substantially from the experiments, which 
cannot be absorbed by the ambiguity of the hidden-gauge coupling 
\(g\) whose value is determined either by 
\(\Gamma(\rho\to 2\pi)\) or by \(\Gamma(\rho\to e^+ e^-)\). 
In order to avoid this ambiguity, let us take some expressions 
cancelling \(g\), i.e. , \(\Gamma(\rho\to\pi\gamma)/
\Gamma(\rho\to 2\pi)\) and \(\Gamma(\omega\to \pi \gamma)/
\Gamma(\rho\to 2\pi)\). 
Then we find that predictions of the original 
\({\cal L}_{{\rm FKTUY\/}}\) and Bramon et al.[8] still do not agree 
with the experiments. 
These Lagrangians without isospin breaking terms yield 
\begin{eqnarray}
\frac{\Gamma(\rho^0 \to \pi^0 \gamma)}{\Gamma(\rho\to 2\pi)}
 &=& \alpha M_{\rho}^2 p_{\rho\to\pi\gamma}^3/16\pi^3 f_{\pi}^2 
 p_{\rho\to\pi\pi}^3 \; ,  \\ 
 &=& 5.7\times 10^{-4} \; [\mbox{from (i) and (ii)}], \nonumber \\
 & & [ (6.6 \pm 0.6) \times 10^{-4} \; \mbox{(iii) Ours } ] \; , 
\nonumber \\
& &  [ (7.9\pm 2.0) \times 10^{-4} \; \mbox{ exp.}  ] \; , 
\nonumber \\
\frac{\Gamma(\rho^{\pm}\to \pi^{\pm} \gamma)}{\Gamma(\rho\to 2\pi)}
 &=& \alpha M_{\rho}^2 p_{\rho\to\pi\gamma}^3/16\pi^3 f_{\pi}^2 
 p_{\rho\to\pi\pi}^3 \; ,  \\ 
 &=& 5.7\times 10^{-4} \; [\mbox{from (i) and (ii)}], \nonumber \\
 & & [ (4.5 \pm 0.5) \times 10^{-4} \; \mbox{(iii) Ours }] \; , 
\nonumber \\
 & & [ (4.5\pm 0.5) \times 10^{-4} \; \mbox{ exp.} ]\;,\nonumber \\
\frac{\Gamma(\omega\to \pi^0 \gamma)}{\Gamma(\rho\to 2\pi)} &=& 
9\alpha M_{\rho}^2 p_{\rho\to\pi\gamma}^3/16\pi^3 f_{\pi}^2 
p_{\rho\to\pi\pi}^3 \; ,  \\
&=& 5.4\times 10^{-3} \; [\mbox{from (i) and (ii)}], \nonumber \\ 
& & [(4.9\pm 0.2)\times 10^{-3} \; \mbox{ (iii) Ours }] \; , 
\nonumber \\
& & [(4.7\pm 0.4)\times 10^{-3} \; \mbox{ exp. } ] \; . \nonumber 
\end{eqnarray}

Finally, we pay attention to 
\(\Gamma(\eta \: (\eta') \: \to \pi^+\pi^-\gamma)\), which are 
given by 
\begin{eqnarray}
\Gamma(\eta\to \pi^+\pi^- \gamma) &=& 
\frac{3 g^2 \alpha}{16\pi^6 f_{\eta}^2 M_{\eta}}
\int dE_+ dE_- [\mbox{\boldmath $p$}_+^2 \mbox{\boldmath $p$}_-^2 - 
(\mbox{\boldmath $p$}_+ \cdot \mbox{\boldmath $p$}_-)^2] \times 
\nonumber \\
 & & \hspace{2.2cm} 
\left ( \frac{1+4/3\epsilon_1+2/3\epsilon_2}{(p_+ +  p_-)^2 - 
 M_{\rho}^2}  + \frac{1+4\epsilon_1+2\epsilon_2}
{3M_{\rho}^2} \right )^2 \; , \label{4-9} \\
\Gamma(\eta'\to \pi^+\pi^- \gamma) &=& 
\frac{3 g^2 \alpha}{32\pi^6 f_{\eta'}^2 M_{\eta'} }
\int dE_+ dE_- [\mbox{\boldmath $p$}_+^2 \mbox{\boldmath $p$}_-^2 - 
(\mbox{\boldmath $p$}_+ \cdot \mbox{\boldmath $p$}_-)^2] \times 
\nonumber \\
 & & \hspace{1.4cm} 
\left ( \frac{1+4/3\epsilon_1+2/3\epsilon_2}{(p_+ +  p_-)^2 +
iM_{\rho} \Gamma_{\rho}- M_{\rho}^2} + 
\frac{1+4\epsilon_1+2\epsilon_2}{3M_{\rho}^2} \right )^2  , 
\label{4-10} \\
\Gamma_{\rho} &=& \Gamma(\rho\to 2\pi) \cdot \left( 
\frac{(p_+ + p_-)^2-4M_{\pi}^2}{M_{\rho}^2-4M_{\pi}^2} 
\right) ^{3/2} 
\theta(\: (p_+ + p_-)^2-4M_{\pi}^2 \: ) \; , \nonumber 
\end{eqnarray}
where $p_{\pm}$ are pion momenta and 
we express \(\rho\)-meson propagater in the process 
\(\eta' \to \rho^0 \gamma \to \pi^+ \pi^- \gamma \) 
by using the \(\rho\)-meson decay width \(\Gamma_{\rho}\). 

One might wonder why the prediction of 
\(\Gamma(\eta (\eta') \to \pi^+ \pi^- \gamma)\) by our model is 
different from the prediction by the original 
\({\cal L}_{\rm FKTUY\/}\) in Table I, 
because each model does not change the low-energy theorem. 
The answer to the question is 
that $p_{\pm}$ in \(\rho\)-meson propagater are not soft momenta 
in the decay of \(\eta (\eta') \to \pi^+ \pi^- \gamma \). 
If we take the low-energy limit in (\ref{4-9}) and (\ref{4-10}), 
we find that isospin/$SU(3)$ breaking effects are cancelled, 
and each model is actually consistent with the low-energy theorem.

\section{The conversion decays of photon into a lepton pair}
\my

In this section, we make the analysis for the decays of 
$V \to P l^+ l^-$. Each decay width is given by 
\begin{equation}
\Gamma(V \to P l^+ l^- ) = 
\frac{\alpha^2}{48\pi M_V}
\int dx_+ dx_- \; T \times \left( \sum_{V'} 
\frac{g_{VV'P}}{(x_+ + x_- -1+\mu_P - \mu_{V'})} 
\frac{g_{V'}}{(x_+ + x_- -1+\mu_P)} \right)^2, 
\end{equation}
\begin{eqnarray*}
T & \equiv & (1-\mu_P-x_+)^2 \cdot (x_+ + x_- -1 +\mu_P) + 
(1-\mu_P-x_-)^2 \cdot (x_+ + x_- -1 +\mu_P) \\ 
 & & - 2 \mu_P (x_+ + x_- -1 +\mu_P)^2 
 -2 \mu_l \left[ 4\mu_P (x_+ + x_- -1 +\mu_P) -
(2-2\mu_P-x_+ - x_-)^2 \right] , \\
x_{\pm} & \equiv & \frac{2E_{l^{\pm}}}{M_V}, \quad 
\mu_{V',P,l}  \equiv  \frac{M_{V',P,l}^2}{M_V^2}.
\end{eqnarray*}
Then we obtain Table II. 

In conversion decays, we can also show importance of 
the isospin breaking effects on \(\omega \to \pi^0 e^+ e^-\). 
As in the previous section, we take an expression eliminating 
the hidden gauge coupling $g$:
\begin{eqnarray}
\frac{\Gamma(\omega \to \pi^0 e^+ e^-)}{\Gamma(\rho \to \pi \pi)} &=&
4.91 \times 10^{-5} 
\mbox{ [original ${\cal L}_{\rm FKTUY\/}$]}, 
\nonumber \\
 & & [(4.42 \pm 0.13) \times 10^{-5} \mbox{ Ours}], \nonumber \\
 & & [(3.29 \pm 1.06) \times 10^{-5} \mbox{ exp.}]. \nonumber 
\end{eqnarray}
Our model reproduces successfully the experimental value of 
$\Gamma(\omega \to \pi^0 e^+ e^-)\) without such a trick as in 
the lattice results of Crisafulli et al.[1], 
who actually made ansatz of linearized expression 
in \(p^2\) of the lepton momentum 
for VMD form factor, assuming \(p^2\) is small. 
We disagree with the lattice results because \(p\) can be the same 
order as \(\omega\) mass(=782MeV), which is not small.

\section{Hadronic Anomalous Decays}
\my

In this section, we consider hadronic anomalous decays such as 
\(\Gamma(\omega\to 3\pi)\). 
In the same way as in the previous section, we obtained Table III. 

In Table III, we took \(K^* K \pi\)-coupling 
as \(1.05 \times g_{K^* K \pi}\), which is given by 
(\ref{HLS}), considering 
\(\Gamma(K^{*\pm} \to (K \pi)^{\pm})=49.8 \pm 0.8\) MeV, 
\(\Gamma(K^{*0} \to (K \pi)^0)=50.5 \pm 0.6\) MeV. 

In Table III it is again suggested 
that isospin breaking terms are very important. 
As in the previous section, 
let us take an expression 
cancelling \(g\), i.e. , \(\Gamma(\omega\to 3\pi)/
\Gamma(\rho\to 2\pi)^3\). 
Then we find that the prediction of the original 
\({\cal L}_{{\rm FKTUY\/}}\) is again different 
from the experiments: 
\begin{eqnarray}
\frac{\Gamma(\omega\to 3\pi)}{\Gamma(\rho\to 2\pi)^3} &=& 
\frac{81M_{\omega} M_{\rho}^6 (1+\epsilon_1+\epsilon_2)^2}
{256\pi^4 f_{\pi}^2 p_{\rho\to\pi\pi}^9 }\int dE_+ dE_- 
[\mbox{\boldmath $p$}_+^2 \mbox{\boldmath $p$}_-^2 - 
(\mbox{\boldmath $p$}_+ \cdot \mbox{\boldmath $p$}_-)^2]\times 
\nonumber \\
 & & \hspace{-1cm} \left( \frac{1}{(p_0 + p_+)^2 + M_{\rho}^2}+
 \frac{1}{(p_+ +  p_-)^2+ M_{\rho}^2}+
 \frac{1}{(p_- + p_0)^2+M_{\rho}^2} \right)^2 \\
\nonumber \\
 &=& 2.38\times 10^{-6} \; {\rm MeV\/}^{-2} 
\mbox{ [ original \({\cal L}_{\rm FKTUY\/}\) with }
\epsilon_1 = \epsilon_2 = 0 \; ], \nonumber \\
 & & [ \; (2.17\pm 0.07)\times 10^{-6} \; {\rm MeV\/}^{-2} \; 
\mbox{ Ours } ] \; , \nonumber \\
 & & [ \; (2.16\pm 0.09)\times 10^{-6} \; {\rm MeV\/}^{-2} \; 
\mbox{ exp. } ] \; . \nonumber 
\end{eqnarray}

Although only the upper bounds of \(\Gamma(\rho\to 3\pi)\) and 
\(\Gamma(K^*\to K\pi\pi)\) are available now, 
it is interesting that their value will be determined 
by the experiments in future.

\section{Summary and Discussions}
\my

By introducing isospin/\(SU(3)\)-broken 
\(\Delta{\cal L}_{VVP}^{a,b}\) with a few parameters, 
we have shown that decay widths of anomalous processes can 
be reproduced consistently with all the experimental data. 

We now discuss the origin of the isospin/$SU(3)$ 
breaking parameters. From (\ref{4-4}), we obtain 
the isospin/\(SU(3)\)-breaking parameters as follows:
\begin{eqnarray}
\epsilon' &=& \left( 
\begin{array}{ccc}
-\epsilon'_1 &  &  \\
 & -0.0292-\epsilon'_1 &  \\
 &  & 0.0366-\epsilon'_1
\end{array}  
\right) , \label{6-1} \\
\epsilon &=& \left( 
\begin{array}{ccc}
-0.0174+\epsilon'_1 &  &  \\
 & 0.0047+\epsilon'_1 &  \\
 &  & 0.1339+\epsilon'_1
\end{array}  
\right), \label{6-2}
\end{eqnarray}
where \(\epsilon'_1\) is an arbitrary parameter.
If we adopt quark mass matrix as isospin/$SU(3)$-breaking terms 
in a usual manner, \(\epsilon'\) of (\ref{6-1}) must be 
proportional to \(\epsilon\) of (\ref{6-2}). 
On this condition, however, we find easily that 
this strategy is absolutely inconsistent, 
even if we consider the error bar of these parameters. 
Thus it seems to need further discussions on 
the origin of these parameters. 
We may suggest that these parameters are deduced from loop effects 
of vector mesons to anomalous \(VVP\)-couplings. In this case, 
we inevitably need to consider the effects of \((VP^3)\)-terms from 
\(\Delta{\cal L}_{1,2,5} \). 
However, their contributions seem to be very small compared with the 
contributions from \(\Delta{\cal L}_{VVP}^{a,b}\), 
because our prediction of \(\Gamma(\omega\to 3\pi)\) is already 
consistent with the experimental value. 

Our predictions for future experiments given in 
Table I, Table II and Table III, are summarized in Table IV. 

We expect that the decay data for pseudoscalar mesons and 
vector mesons, such as 
\(\phi\to\eta'\gamma,\rho^0\to\pi^0\gamma,\phi \to \pi^0 e^+ e^- \) 
and so on, will be obtained with good accuracy in 
the DA\(\Phi\)NE \(\phi\)-factory. 

\section*{ \hspace{4cm} Acknowledgements}

The author is very grateful to K. Yamawaki for suggesting 
this subject, helpful discussions and also for careful reading 
the manuscript. 
Thanks are also due to the members of the theory group 
at Nagoya University for discussions and encouragement.

\newpage

\footnotesize
\begin{table}[t]
\caption[table]{Radiative Decay Width of Vector Mesons \par 
(i) Values of original \({\cal L}_{{\rm FKTUY\/}}\) [7] \\
(ii) Values of the \(SU(3)\)-broken model by Bramon et al.[8]  
(\(\epsilon_3 = -0.1 \pm 0.03\)).   \\
(iii) Values of our model by \(\chi^2\)-fitting. 
The parameters are \\
$\epsilon_1=-0.0174 \pm 0.0100$, 
$\epsilon_2=-0.0246 \pm 0.0114$, $\epsilon_3=-0.0974 \pm 0.0103$, \\
$\epsilon'_1 - \epsilon'_2 = -0.0292 \pm 0.0015 $ and 
$\epsilon'_1 - \epsilon'_3 = 0.0366 \pm 0.0028 $. }
\begin{tabular}{l|rrr||r} \hline
{Decay Mode} & {(i) \({\cal L}_{{\rm FKTUY\/}}\) [7]} & 
{(ii) Bramon's [8] } & {(iii) Ours \quad} \qquad & {exp.[2]} \qquad 
\\ \hline \hline
\(\Gamma(\rho^0\to\pi^0\gamma) \) &  \(86.2\pm 0.8{\rm keV\/}\) 
&  \(86.2\pm 0.8{\rm keV\/}\) &  
\(101 \pm 9 {\rm keV\/}\) &  \(121\pm 31{\rm keV\/}\) \\
\(\Gamma(\rho^{\pm}\to\pi^{\pm}\gamma) \) & \(85.6\pm 0.8{\rm keV\/}\) 
& \(85.6\pm 0.8{\rm keV\/}\) & \(68.1\pm 7.1{\rm keV\/}\) 
& \(68\pm 7{\rm keV\/}\) \\
\(\Gamma(\omega\to\pi^0\gamma ) \) &\(815\pm 8{\rm keV\/}\) 
& \(815\pm 8{\rm keV\/}\) &  \(734\pm 21{\rm keV\/}\)  
& \(717\pm 43{\rm keV\/}\) \\
\(\Gamma(\omega\to\eta\gamma) \) & \(6.68\pm 0.59{\rm keV\/}\) 
& \(5.6\pm 0.6{\rm keV\/}\) & \(4.17\pm 0.77{\rm keV\/}\) 
& \(7.00\pm 1.77{\rm keV\/}\) \\
\(\Gamma(\rho^0\to\eta\gamma) \) & \(52.4\pm 4.6{\rm keV\/}\) 
& \(52.4\pm 4.6{\rm keV\/}\) & 
\(49.5\pm 4.9{\rm keV\/}\) & \(57.5\pm 10.6 {\rm keV\/}\) \\
\(\Gamma(\phi\to\eta\gamma) \) & \(80.7 \pm 7.1{\rm keV\/}\) 
& \(57 \pm 9{\rm keV\/}\) & \(58.0\pm 7.6{\rm keV\/}\) 
& \(56.9\pm 2.9{\rm keV\/}\) \\
\(\Gamma(K^{*\pm}\to K^{\pm}\gamma) \) & \(32.8\pm 0.9{\rm keV\/}\) 
& \(47\pm 5{\rm keV\/}\) & \(50.0 \pm 3.9{\rm keV\/}\) 
& \(50\pm 5{\rm keV\/}\) \\ 
\(\Gamma(\bar{K^{*0}}\to\bar{K^0}\gamma) \) &\(132\pm 4{\rm keV\/}\) 
& \(107\pm 15{\rm keV\/}\)& \(102 \pm 5{\rm keV\/}\) 
& \(117\pm 10{\rm keV\/}\) \\
\(\Gamma(\phi\to\pi^0\gamma) \) & \(----\) 
& \(6.76\pm 0.34{\rm keV\/}\) & \(6.09\pm 0.39{\rm keV\/}\) 
& \(5.80\pm 0.58{\rm keV\/}\) \\
\(\Gamma(\eta'\to\rho^0\gamma) \) & \(61.7\pm 2.7{\rm keV\/}\) 
& \(61.7\pm 2.7{\rm keV\/}\) & \(58.3\pm 3.3{\rm keV\/}\) 
& \(61\pm 5{\rm keV\/}\) \\
\(\Gamma( \eta'\to\omega \gamma) \) & \(5.74\pm 0.25{\rm keV\/}\) 
& \(7.86 \pm 0.34{\rm keV\/}\) & \(6.27\pm 0.61 {\rm keV\/}\) 
& \(6.1\pm 0.8{\rm keV\/}\) \\
\(\Gamma(\phi\to\eta'\gamma) \) & \(0.827\pm 0.036{\rm keV\/}\) 
& \(0.5\pm 0.1{\rm keV\/}\) & 
\(0.508\pm 0.035{\rm keV\/}\) & \(<1.84{\rm keV\/}\) \\ \hline \hline
\(\Gamma(\pi^0\to 2\gamma) \) & \(7.70{\rm eV\/}\) & 
\(7.70{\rm eV\/}\) & \(7.70{\rm eV\/}\) & \(7.7 \pm 0.6 {\rm eV\/}\) \\
\(\Gamma(\eta\to 2\gamma) \) & \(0.46\pm 0.04{\rm keV\/}\) 
& \(0.51\pm 0.04{\rm keV\/}\) & \(0.46\pm 0.04{\rm keV\/}\) 
& \(0.46\pm 0.04{\rm keV\/}\) \\
\(\Gamma(\eta'\to 2\gamma) \) & \(4.26\pm 0.19{\rm keV\/}\) 
& \(3.6\pm 0.2{\rm keV\/}\) & \(4.26\pm 0.19{\rm keV\/}\) 
& \(4.26\pm 0.19{\rm keV\/}\) \\ \hline \hline
\(\Gamma(\eta\to \pi^+\pi^- \gamma)\) &\(0.0660\pm 0.0053{\rm keV\/}\) 
& \(0.0660\pm 0.0053{\rm keV\/}\) 
& \(0.0641\pm 0.0054{\rm keV\/}\) & \(0.0586\pm 0.0057{\rm keV\/}\) \\
\(\Gamma(\eta'\to \pi^+ \pi^- \gamma) \) & \(53.0\pm 2.2{\rm keV\/}\) 
& \(53.0\pm 2.2{\rm keV\/}\) 
& \(49.5\pm 2.1{\rm keV\/}\) & \(56.1\pm 6.4{\rm keV\/}\) \\ \hline 
\hline
\end{tabular}
\end{table}
\normalsize

\hspace{1cm} 

\newpage

\footnotesize

\begin{table}[t]
\caption[table]{Decay Width of \(V \to P l^+ l^-\) \par  
In our model, the parameters are \\
$\epsilon_1=-0.0174 \pm 0.0100$, 
$\epsilon_2=-0.0246 \pm 0.0114$, $\epsilon_3=-0.0974 \pm 0.0103$, \\
$\epsilon'_1 - \epsilon'_2 = -0.0292 \pm 0.0015 $ and 
$\epsilon'_1 - \epsilon'_3 = 0.0366 \pm 0.0028 $. }
\begin{tabular}{l|rr||r} \hline
{\quad Decay Mode} & { \({\cal L}_{{\rm FKTUY\/}} \) [7] \qquad} & 
{ Ours \qquad \quad} & {exp.[2] \quad }  \\ \hline \hline
\(\Gamma(\rho^0\to\pi^0 e^+ e^-) \) & \(0.778 \pm 0.007{\rm keV\/}\) 
& \(0.914 \pm 0.079 {\rm keV\/}\) &  \(-----\) \\
\(\Gamma(\rho^0\to\pi^0 \mu^+ \mu^-) \) 
& \(0.0734 \pm 0.0007{\rm keV\/}\) & \(0.0863 \pm 0.0075 {\rm keV\/}\) 
&  \(-----\) \\
\(\Gamma(\rho^{\pm}\to\pi^{\pm}e^+ e^-) \) 
& \(0.771\pm 0.007 {\rm keV\/}\) & \(0.614 \pm 0.065 {\rm keV\/}\) 
&  \(-----\)\\
\(\Gamma(\rho^{\pm}\to\pi^{\pm} \mu^+ \mu^-) \) 
& \(0.0719\pm 0.0007 {\rm keV\/}\) & \(0.0572 \pm 0.0060 {\rm keV\/}\) 
&  \(-----\)\\
\(\Gamma(\omega\to\pi^0 e^+ e^-) \) 
& \(7.42\pm 0.07{\rm keV\/}\) & \(6.68 \pm 0.22{\rm keV\/}\) 
&  \(4.97 \pm 1.60{\rm keV\/} \) \\
\(\Gamma(\omega\to\pi^0 \mu^+ \mu^-) \) 
& \(0.745\pm 0.007{\rm keV\/}\) & \(0.670 \pm 0.022{\rm keV\/}\) 
&  \(0.809 \pm 0.194{\rm keV\/} \) \\
\(\Gamma(\rho^0 \to \eta e^+ e^-) \) 
& \(0.377 \pm 0.031{\rm keV\/}\) & \(0.301 \pm 0.039 {\rm keV\/}\) 
&  \(-----\)\\
\(\Gamma(\rho^0 \to \eta \mu^+ \mu^-) \) 
& \((1.40 \pm 0.11) \times 10^{-5}{\rm keV\/}\) 
& \((1.12 \pm 0.15) \times 10^{-5}{\rm keV\/}\) 
&  \(-----\)\\
\(\Gamma(\omega\to\eta e^+ e^-) \) 
& \(0.0487 \pm 0.0039{\rm keV\/}\) & \(0.0304 \pm 0.0043{\rm keV\/}\) 
&  \(-----\)\\
\(\Gamma(\omega\to\eta \mu^+ \mu^-) \) 
& \((1.36 \pm 0.11) \times 10^{-5}{\rm keV\/}\) 
& \((0.850 \pm 0.012) \times 10^{-5}{\rm keV\/}\) 
&  \(-----\)\\
\(\Gamma(\phi\to\eta e^+ e^-) \) 
& \(0.677 \pm 0.055 {\rm keV\/}\) & \(0.486\pm 0.046{\rm keV\/}\) 
& \(0.576 
\mbox{\tiny
$\begin{array}{c}
+0.354 \\
-0.266
\end{array} 
$}
{\rm keV\/}\) \\
\(\Gamma(\phi\to\eta \mu^+ \mu^-) \) 
& \(0.0325 \pm 0.0026 {\rm keV\/}\) & \(0.0233\pm 0.0022{\rm keV\/}\) 
& \(-----\) \\
\(\Gamma(K^{*\pm}\to K^{\pm} e^+ e^-) \) 
& \(0.272 \pm 0.007{\rm keV\/}\) & \(0.414\pm 0.032{\rm keV\/}\) 
& \(-----\) \\
\(\Gamma(K^{*\pm}\to K^{\pm} \mu^+ \mu^-) \) 
& \(0.00962 \pm 0.00026{\rm keV\/}\) & \(0.0141\pm 0.0011{\rm keV\/}\) 
& \(-----\) \\
\(\Gamma(\bar{K^{*0}}\to\bar{K^0} e^+ e^-) \) 
& \(4.32 \pm 0.12{\rm keV\/}\) & \(3.34 \pm 0.15{\rm keV\/}\) 
& \(-----\) \\
\(\Gamma(\bar{K^{*0}}\to\bar{K^0} \mu^+ \mu^-) \) 
& \(0.131 \pm 0.003{\rm keV\/}\) & \(0.102 \pm 0.004{\rm keV\/}\) 
& \(-----\) \\
\(\Gamma(\phi\to\pi^0 e^+ e^-) \) 
& \(-----\) & \(0.0640 \pm 0.0042{\rm keV\/}\) 
& \(< 0.532 {\rm keV\/} \) \\
\(\Gamma(\phi\to\pi^0 \mu^+ \mu^-) \) 
& \(-----\) & \(0.0137 \pm 0.0009{\rm keV\/}\) 
& \(----- \) \\
\(\Gamma(\eta'\to\rho^0 e^+ e^-) \) 
& \(0.428 \pm 0.019{\rm keV\/}\) & \(0.341 \pm 0.024{\rm keV\/}\) 
& \(-----\) \\
\(\Gamma( \eta'\to\omega e^+ e^-) \) 
& \(0.0392 \pm 0.0017 {\rm keV\/}\) & \(0.0245 \pm 0.0031{\rm keV\/}\) 
& \(-----\) \\
\(\Gamma(\phi\to\eta' e^+ e^-) \) 
& \(0.00407 \pm 0.00018{\rm keV\/}\) & \(0.00262 \pm 0.00018{\rm keV\/}\) 
& \(-----\) \\ \hline \hline
\(\Gamma(\pi^0 \to \gamma e^+ e^- ) \) 
& \(0.0920 {\rm eV\/}\) 
& \(0.0920 \pm 0.0001 {\rm eV\/}\) 
& \(0.0922 \pm 0.0076 {\rm eV\/}\) \\ 
\(\Gamma(\eta \to \gamma e^+ e^- ) \) 
& \(7.79 \pm 0.62{\rm eV\/}\) & \(7.78 \pm 0.62{\rm eV\/}\) 
& \(6.00 \pm 1.54 {\rm eV\/} \) \\ 
\(\Gamma(\eta \to \gamma \mu^+ \mu^- ) \) 
& \(0.373 \pm 0.030 {\rm eV\/}\) & \(0.370 \pm 0.031{\rm eV\/}\) 
& \(0.372 \pm 0.059 {\rm eV\/} \) \\ 
\(\Gamma(\eta'\to \gamma e^+ e^- ) \) 
& \(0.0859 \pm 0.0036 {\rm keV\/}\) & \(0.0852 \pm 0.0037{\rm keV\/}\) 
& \(-----\) \\ 
\(\Gamma(\eta'\to \gamma \mu^+ \mu^- ) \) 
& \(0.0192 \pm 0.0008 {\rm keV\/}\) & \(0.0183 \pm 0.0014{\rm keV\/}\) 
& \(0.0209 \pm 0.0055 {\rm keV\/} \) \\ \hline \hline
\end{tabular}
\end{table}
\normalsize

\hspace{1cm}

\newpage

\footnotesize

\begin{table}[h]
\caption[table]{Hadronic Decay Width of Vector Mesons \par  
In our model, the parameters are \\
$\epsilon_1=-0.0174 \pm 0.0100$, 
$\epsilon_2=-0.0246 \pm 0.0114$, $\epsilon_3=-0.0974 \pm 0.0103$, \\
$\epsilon'_1 - \epsilon'_2 = -0.0292 \pm 0.0015 $ and 
$\epsilon'_1 - \epsilon'_3 = 0.0366 \pm 0.0028 $. }
\begin{tabular}{l|rr||r} \hline
{\quad Decay Mode} & { \({\cal L}_{{\rm FKTUY\/}} \) [7] \quad}  & 
{ Ours \qquad }  & {exp. [2] \quad}  \\ \hline \hline
\(\Gamma(\omega\to \pi^0 \pi^+ \pi^-) \) & \(8.18\pm 0.23{\rm MeV\/}\) 
& \(7.51\pm 0.33{\rm MeV\/}\) 
& \(7.49\pm 0.12{\rm MeV\/}\) \\ \hline
\(\Gamma(\rho^0 \to \pi^0 \pi^+ \pi^-) \) & \(----\)  
& \(8.12\pm 2.35 {\rm keV\/}\) & \(<18{\rm keV\/}\) \\
\(\Gamma(\rho^{\pm} \to \pi^{\pm} \pi^0 \pi^0) \)  
& \(----\) & \(2.11\pm 0.43{\rm keV\/}\) & \(-----\) \\
\(\Gamma(\rho^{\pm} \to \pi^{\pm} \pi^+ \pi^-) \)  
& \(----\) & \(0.141 \pm 0.071{\rm keV\/}\) & \(-----\) \\
\(\Gamma(K^{*-} \to \bar{K^0} \pi^0 \pi^-) \) 
& \(17.9\pm 0.7{\rm keV\/}\)  
& \(11.5\pm 0.5{\rm keV\/}\) & \(<35{\rm keV\/}\) \\
\(\Gamma(K^{*-} \to K^- \pi^+ \pi^-) \) 
& \(8.65\pm 0.33{\rm keV\/}\) 
& \(5.83\pm 0.23{\rm keV\/}\) & \(<40{\rm keV\/}\) \\
\(\Gamma(K^{*-} \to K^- \pi^0 \pi^0) \) 
& \(1.11 \pm 0.04{\rm keV\/}\)  
& \(0.593 \pm 0.025{\rm keV\/}\) & \(-----\) \\
\(\Gamma(\bar{K^{*0}} \to K^- \pi^0 \pi^+) \) 
& \(23.2 \pm 0.9{\rm keV\/}\) 
& \(15.9\pm 0.6{\rm keV\/}\) & \(-----\) \\
\(\Gamma(\bar{K^{*0}} \to \bar{K^0} \pi^- \pi^+) \) 
& \(9.04\pm 0.34{\rm keV\/}\) 
& \(5.94\pm 0.23{\rm keV\/}\) & \(< 35{\rm keV\/}\) \\
\(\Gamma(\bar{K^{*0}} \to \bar{K^0} \pi^0 \pi^0) \) 
& \(1.11\pm 0.05{\rm keV\/}\)
& \(0.507\pm 0.022{\rm keV\/}\) & \(-----\) \\
\hline \hline
\end{tabular}
\end{table}
\normalsize

\hspace{1cm}

\newpage

\footnotesize
\begin{table}[h]
\caption[table]{The List of Our Predictions}
\begin{center}
\begin{tabular}{|r|lc|} \hline
{ Decay Mode \quad} & {Decay Width [keV]} & { B.R. } \\
\hline \hline
\(\Gamma(\rho^0\to\pi^0\gamma)\) & \(101 \pm 9 \) & 
\((6.68 \pm 0.58)\times 10^{-4}\) \\
\(\Gamma(\phi\to\eta'\gamma)\) &  \(0.508\pm 0.035\) & 
\((1.15 \pm 0.08) \times 10^{-4}\)  \\ \hline
\(\Gamma(\rho^0\to\pi^0 e^+ e^-) \) 
& \(0.914 \pm 0.079 \) &  \((6.08 \pm 0.53) \times 10^{-6}\) \\
\(\Gamma(\rho^0\to\pi^0 \mu^+ \mu^-) \) 
& \(0.0863 \pm 0.0075 \) 
&  \((5.71\pm 0.50) \times 10^{-7}\) \\
\(\Gamma(\rho^{\pm}\to\pi^{\pm}e^+ e^-) \) 
& \(0.614 \pm 0.065 \) 
&  \((4.61 \pm 0.43) \times 10^{-6} \) \\
\(\Gamma(\rho^{\pm}\to\pi^{\pm} \mu^+ \mu^-) \) 
& \(0.0572 \pm 0.0060 \) 
& \((3.78\pm 0.40) \times 10^{-7} \) \\
\(\Gamma(\rho^0 \to \eta e^+ e^-) \) 
& \(0.301 \pm 0.039 \) 
& \((1.99 \pm 0.26) \times 10^{-6} \) \\
\(\Gamma(\omega\to\eta e^+ e^-) \) 
& \(0.0304 \pm 0.0043 \) 
& \((3.61 \pm 0.51) \times 10^{-6} \) \\
\(\Gamma(\phi\to\eta e^+ e^-) \) 
& \(0.486\pm 0.046 \) 
& \(1.10 \pm 0.10) \times 10^{-4} \) \\
\(\Gamma(\phi\to\eta \mu^+ \mu^-) \) 
& \(0.0233\pm 0.0022 \) 
& \((5.26 \pm 0.50) \times 10^{-6} \) \\
\(\Gamma(K^{*\pm}\to K^{\pm} e^+ e^-) \) 
& \(0.414\pm 0.032 \) 
& \((8.31 \pm 0.66) \times 10^{-6} \) \\
\(\Gamma(K^{*\pm}\to K^{\pm} \mu^+ \mu^-) \) 
& \(0.0141\pm 0.0011 \) 
& \((2.83 \pm 0.23) \times 10^{-7} \) \\
\(\Gamma(\bar{K^{*0}}\to\bar{K^0} e^+ e^-) \) 
& \(3.34 \pm 0.15 \) 
& \((6.61 \pm 0.31) \times 10^{-5} \) \\
\(\Gamma(\bar{K^{*0}}\to\bar{K^0} \mu^+ \mu^-) \) 
& \(0.102 \pm 0.004 \) 
& \((2.02 \pm 0.08) \times 10^{-6} \) \\
\(\Gamma(\phi\to\pi^0 e^+ e^-) \) 
& \(0.0640 \pm 0.0042 \) 
& \((1.44 \pm 0.10) \times 10^{-5} \) \\
\(\Gamma(\phi\to\pi^0 \mu^+ \mu^-) \) 
& \(0.0137 \pm 0.0009 \) 
& \((3.09 \pm 0.21) \times 10^{-6} \) \\
\(\Gamma(\eta'\to\rho^0 e^+ e^-) \) 
& \(0.341 \pm 0.024 \) 
& \(1.70 \pm 0.18) \times 10^{-3} \) \\
\(\Gamma( \eta'\to\omega e^+ e^-) \) 
& \(0.0245 \pm 0.0031 \) 
& \((1.22 \pm 0.18) \times 10^{-4} \) \\
\(\Gamma(\phi\to\eta' e^+ e^-) \) 
& \(0.00262 \pm 0.00018 \) 
& \((5.91 \pm 0.41) \times 10^{-7} \) \\
\(\Gamma(\eta' \to \gamma e^+ e^-) \) & \(0.0852 \pm 0.0037\) 
& \((4.24 \pm 0.38) \times 10^{-4} \) \\ \hline
\(\Gamma(\rho^0 \to \pi^0 \pi^+ \pi^-)\) &
\(8.12\pm 2.35\) & \((5.37\pm 1.56)\times 10^{-5}\)  \\
\(\Gamma(\rho^{\pm} \to \pi^{\pm} \pi^0 \pi^0)\) & 
\(2.11 \pm 0.43\) & \((1.40 \pm 0.28) \times 10^{-5}\)  \\
\(\Gamma(\rho^{\pm} \to \pi^{\pm} \pi^+ \pi^-)\) & 
\(0.141 \pm 0.071\) & \((9.32 \pm 4.71) \times 10^{-7}\)  \\
\(\Gamma(K^{*-} \to \bar{K^0} \pi^0 \pi^-)\) & 
\(11.5\pm 0.5\) & \((2.31\pm 0.10) \times 10^{-4}\)  \\
\(\Gamma(K^{*-} \to K^- \pi^+ \pi^-)\) & 
\(5.83\pm 0.23\) & \((1.17 \pm 0.05) \times 10^{-4}\)  \\
\(\Gamma(K^{*-} \to K^- \pi^0 \pi^0)\) &  
\(0.593 \pm 0.025\) & \((1.19 \pm 0.05) \times 10^{-5}\)  \\
\(\Gamma(\bar{K^{*0}} \to K^- \pi^0 \pi^+)\) & 
\(15.9\pm 0.6\) & \((3.15 \pm 0.13) \times 10^{-4}\)   \\
\(\Gamma(\bar{K^{*0}} \to \bar{K^0} \pi^- \pi^+)\) & 
\(5.94 \pm 0.23\) & \((1.18 \pm 0.05) \times 10^{-4}\)   \\
\(\Gamma(\bar{K^{*0}} \to \bar{K^0} \pi^0 \pi^0)\) & 
\(0.507\pm 0.022\) & \((1.00 \pm 0.04) \times 10^{-5}\)  \\ \hline 
\end{tabular}
\end{center}
\end{table}
\normalsize


\begin{thebibliography}{99}
\bibitem{1} A. Bramon, A. Grau and G. Pancheri, 
 in ``Second DA\(\Phi\)NE Physics Handbook'',
 eds. L. Maiani, G. Pancheri and N. Paver (INFN, Franzini, 1995),
 p. 477; 
M. Crisafulli, V. Lubicz, {\it idid}, p. 515; 
S. I. Eidelman, {\it ibid}, p. 523; 
J. Lee-Franzini, {\it ibid}, p. 761. 

\bibitem{2} Particle Data Group, Phys. Rev. {\bf D50}, 1173 (1994).

\bibitem{3} J. Wess and B. B. Zumino, Phys. Lett. 
{\bf 37B}, 95 (1971): 
E. Witten, Nucl. Phys. {\bf B223}, 422 (1983).

\bibitem{4} M. Bando, T. Kugo, S. Uehara, K. Yamawaki, and 
T. Yanagida,  Phys. Rev. Lett. {\bf 54}, 1215 (1985).

\bibitem{5} M. Bando, T. Kugo and K. Yamawaki, 
Nucl. Phys. {\bf B259}, 493 (1985).

\bibitem{6} M. Bando, T. Kugo and K. Yamawaki, 
Phys. Rep. {\bf 164}, 217 (1988).

\bibitem{7} T. Fujiwara, T. Kugo, H. Terao, S. Uehara and 
K. Yamawaki, Prog. Theor. Phys. {\bf 73}, 926 (1985).

\bibitem{8} A. Bramon, A. Grau and G. Pancheri, Phys. Lett. 
{\bf B344}, 240 (1995).

\bibitem{9} M. Hashimoto, Nagoya Univ. Preprint DPNU-96-08, 
hep-ph/9602338, to appear in Phys. Lett. {\bf B}.

\bibitem{10} M. Harada and J. Shecheter, Preprint, 
hep-ph/9506473.

\bibitem{11} J. Gasser and H. Leutwyler, Ann. Phys. {\bf 158}, 142 
(1984): Nucl. Phys. {\bf B250}, 465 (1985). 

\bibitem{12} D. Alde et al., Z. Phys. {\bf C61}, 35 (1994).

\bibitem{13} J. Gilman and R. Kauffman, Phys. Rev. {\bf D36}, 2761 
(1987).

\bibitem{14} S. I. Dolinsky et al., Z. Phys. {\bf C42}, 511 (1989).

\end{thebibliography}
\end{document}